\documentclass{ws-ijmpa}
\usepackage[super,compress]{cite}
\usepackage{amsmath}
\usepackage{amssymb}
\usepackage{graphicx}

\let\oldsqrt\sqrt
\def\sqrt{\mathpalette\DHLhksqrt}
\def\DHLhksqrt#1#2{%
\setbox0=\hbox{$#1\oldsqrt{#2\,}$}\dimen0=\ht0
\advance\dimen0-0.2\ht0
\setbox2=\hbox{\vrule height\ht0 depth -\dimen0}%
{\box0\lower0.4pt\box2}}

\begin{document}
\markboth{Katz}{A Quantum Operator for Distance}

%
\catchline{}{}{}{}{}
%

\title{A New Quantum Operator for Distance}

\author{Daniel Katz}

\address{Department of Physics \& Applied Physics\\
University of Massachusetts Lowell\\
1 University Avenue, Lowell, MA 01854, USA\\
daniel\_katz@uml.edu}

\maketitle

\begin{history}
\received{Day Month Year}
\revised{Day Month Year}
\end{history}

\begin{abstract}
We introduce a new semi-relativistic quantum operator for the length of the worldline a particle traces out as it moves. In this article the operator is constructed in a heuristic way and some of its elementary properties are explored. The operator ends up depending in a very complicated way on the potential of the system it is to act on so as a proof of concept we use it to analyze the expected distance traveled by a free Gaussian wavepacket with some initial momentum. It is shown in this case that the distance such a particle travels becomes light-like as its mass vanishes and agrees with the classical result for macroscopic masses. This preliminary result has minor implications for the Weak Equivalence Principle (WEP) in quantum mechanics. In particular it shows that the logical relationship between two formulations of the WEP in classical mechanics extends to quantum mechanics. That our result is qualitatively consistent with the work of others emboldens us to start the task of evaluating the new operator in non-zero potentials. However, we readily acknowledge that the looseness in the definition of our operator means that all of our so-called results are highly speculative. Plans for future work with the new operator are discussed in the last section.

\end{abstract}

\ccode{PACS numbers: 03.65.-w, 04.60.-m}


\section{Introduction}
\label{sec:intro}
It is quite well-known that the simple combination of quantum mechanics/field theory with Einstein's General Relativity (GR) leads to nonsensical predictions. One reason for this may be the incompatibility of the WEP of GR with the type of statements one is allowed to make about quantum particles. While it does not rigorously justify GR - that honor belongs to the somewhat more restrictive Einstein Equivalence Principle -  the WEP is the conceptual foundation on which metric theories of gravity are based. To study quantum gravity many have sought a quantum analog to the WEP, but these efforts are frustrated by the non-classical nature of quantum mechanics. For instance, many versions of the classical WEP refer to an object's trajectory but such a concept does not exist for a quantum particle. To motivate the introduction of a new operator we begin by briefly reviewing the WEP in classical mechanics and some of the ways in which it may (or may not) apply to quantum mechanics.

In classical mechanics there are many different ways of formulating the WEP. At the risk of oversimplifying, we can sort the principles into two broad categories. One category contains those principles which assert that an object's trajectory does not depend on its own properties so long as it is subjected to no non-gravitational forces. The other category says nothing directly about trajectories, but rather identifies inertial mass with gravitational mass for all objects. While it is well-known that these two categories, or modifications thereof, exist there does not seem to be a consensus in the community about what to call them so for this discussion we assign the two categories the names Universality of Free Fall (UFF) and Newton's Equivalence Principle (NEP). Exemplars of each category are, more explicitly,
\begin{quote}
\textbf{UFF:} If an uncharged test body is placed at an initial event in spacetime and is given an initial velocity there, then its subsequent worldline will be independent of its internal structure and composition~\cite{GRsurvey,thorne}.
\end{quote}

\noindent and

\begin{quote}
\textbf{NEP:} In the Newtonian limit the inertial and gravitational masses of a body are equal~\cite{nonequiv}.
\end{quote}

Note that this definition of UFF is slightly more general than its more common formulation which essentially replaces the phrase ``initial event in spacetime" with ``point in a gravitational field." The latter phrasing is used, for instance, in Refs.~\citen{schutz} and~\citen{nonequiv} but we use the former so that UFF applies, albeit trivially, to free particles. In much of the literature and GR textbooks the two formulations of the WEP are simply taken to be equivalent to each other (see, for example, Refs.~\citen{rindler} and~\citen{haugan}). Our definition of the NEP comes directly from a paper by Casola, Liberati and Sonego~\cite{nonequiv} in which they lay out a good deal of variations on the WEP and analyze the logical dependencies among them in the context of classical physics. They show, among other things, that the truth of UFF necessitates the truth of NEP but also that the implication is one-way. That is, UFF $\to$ NEP and yet NEP $\nrightarrow$ UFF in classical mechanics. The main result of the present work is that this inequivalence of equivalence principles extends into the quantum domain (at least for quantum states having classical analogs). Gravity as a force is ``switched off" in this paper so we make no claims yet of testing the $m_i = m_g$ hypothesis itself theoretically. However by demonstrating that the mean length of the worldline an initially well-localized free quantum particle traverses depends on its mass, we show that UFF fails to hold in the quantum theory regardless of NEP. That is, $NEP \nrightarrow UFF$ vacuously in quantum mechanics. One is almost forced to work with expectation values since quantum particles generally lack definite properties, most of all trajectory/worldline. This can be justified by an intuitive appeal to Ehrenfest's theorem, which was the approach taken by Greenberger in the appendix of Ref.~\citen{neutron} to show that WEP is only an approximate symmetry in quantum theory which becomes exact in the limit of large quantum numbers. He points out that the spread of a localized Gaussian wavefunction evolves in a mass-dependent way, but that contributions to observables from this spread vanish in the classical limit. Put another way, low-lying states overlap significantly with each other while semi-classical states are more or less distinct. 

One widely used technique for studying the WEP in quantum mechanics involves making statements about mean arrival times of ensembles of freely falling\footnote{If the particles in such a situation are to me monitored \emph{as} they fall, as opposed to just when they land, then the meaning of the ``free" part of free-fall must be reexamined.} particles. Prominent examples include Refs.~\citen{ali,viola} and \citen{davies} which all conclude that when quantum effects are important WEP violation is observed, usually through explicit mass-dependence of observables, but that the WEP is restored in the classical limit. Actually, Ref.~\citen{viola} found that the mean arrival time of an ensemble of freely falling Gaussian wavepackets is independent of the particle's mass but that higher order observables (\emph{e.g.} the variance of the arrival time) are mass dependent. They also show that wherever mass dependence of observables occurs for the freely falling particle the exact same mass dependence is seen for free particles observed from a uniformly accelerating reference frame. They conclude (in the language of this article) that NEP implies a variant of UFF for quantum mechanical objects. At first blush it looks like Ref.~\citen{viola} is contradicting Ref.~\citen{neutron}, but this is not the case since they are making claims about slightly different formulations of the WEP. Among those who study quantum time of flight there is disagreement about what exactly ``quantum time of flight" means. Some workers~\cite{viola} take the time of flight to be the solution to Ehrenfest's analog of the kinematic equation $\langle z(t)\rangle = \langle z(0)\rangle + (\langle p\rangle/m)t - gt^2/2$ while others~\cite{timeop} take it to be the expectation value of a time operator, defined to be conjugate to the system's Hamiltonian. These two approaches are inequivalent and hardly represent all the various ways in which time is interpreted in quantum mechanics. 

In this article we present a new tool to add to the fray: a quantum operator for special relativistic four-distance. It was originally developed with the intention of analyzing the invariant interval of a particle's motion in a gravitational field, although that is not covered here. The rest of this work is laid out as follows. In section~\ref{sec:der} we motivate the distance operator by quantizing the distance element of Minkowski spacetime. As ``length of worldline" is not a dynamical variable of the classical theory, our quantization scheme cannot follow the canonical plan of replacing Poisson brackets with commutators and inserting factors of $i\hbar$. Instead, we take the non-rigorous route of replacing classical position and momentum variables with their corresponding quantum mechanical operators. The distance operator is time-extended and in section~\ref{sec:measure} we discuss how its expectation values might be obtained experimentally using nondestructive weak measurements. The operator ends up depending in a complicated way on the potential of the system it is to act on, so as a proof of concept we begin by analyzing in section~\ref{sec:free} free particles, both localized Gaussian wavepackets and delocalized planewaves. We will see that the expected distance traveled by a Gaussian depends on the particle's mass but that this dependence vanishes in the classical limit. Having justified the study of the new operator, in section~\ref{sec:fut} we discuss what future work might be done with it. Natural units ($c=\hbar=1$) are used throughout.

\section{Construction of the Operator}
\label{sec:der}
We begin by considering the distance element in four-dimensional Minkowski space parameterized by time,
\begin{equation}
ds = \sqrt{1-\frac{d\vec{x}}{d t}\cdot \frac{d\vec{x}}{d t}}d t.
\label{eq:mink}
\end{equation}
Since the Minkowski spacetime structure is in some sense the ``kernel" of Special Relativity, this starting point is the sense in which the operator of this paper is partially relativistic. The explicit time dependence complicates the evolution of the operator which will result from quantization of this object and we will need the expressions
\begin{equation}
s = \int_0^t d t' \sqrt{1-\frac{d\vec{x}}{d t'}\cdot \frac{d\vec{x}}{d t'}}
\label{eq:stint}
\end{equation}
and
\begin{equation}
\frac{\partial s}{\partial t} =  \sqrt{1-\frac{d\vec{x}}{d t}\cdot \frac{d\vec{x}}{d t}},
\label{eq:ds}
\end{equation}
which follow immediately from~Eq.~\eqref{eq:mink}. The derivative of $s$ in~Eq.~\eqref{eq:ds} is written as a partial in order to distinguish it from the total differential which occurs on the left-hand-side of Ehrenfest's theorem. For simplicity we have chosen $t_0 = 0$ for the initial time in~Eq.~\eqref{eq:stint} or equivalently, if one wishes to think of the integral as indefinite, we choose zero for the integration constant. To quantize $s$ we take the approach of promoting $\vec{x}$ to an operator whose components obey the commutation relations
\begin{equation}
[x_i,p_j] =i \delta_{ij}, \quad i,j=1,2,3,
\end{equation}
where $\delta_{ij}$ is a Kronecker delta, with their conjugates, the components of the momentum operator $\vec{p}$. Now $\vec{x}$ is no longer a curve parameterized by time and representing the trajectory of a particle, but rather a quantum operator whose evolution is determined by Heisenberg's equation. By invoking Heisenberg we are giving up some generality as his equation breaks down before $p \sim m$. Though it is not in very common use, a relativistic analog of Heisenberg's equation does exist~\cite{covheis,covheis1} and one may therefore wonder why we are proceeding without it. The reasons are twofold. First, in this preliminary investigation we wish to keep things simple and adding relativity to the mix tends to do the opposite. Second, using a relativistic equation for the evolution of an operator would oblige us to use relativistic wavefunctions to compute expectation values, which bring with them a whole host of technical and interpretation problems. We elect to defer dealing with these issues. Although we are dealing with only free particles in this paper, for future purposes it will be useful to know the evolution of our distance operator in the presence of a potential depending on at most $\vec{x}$. The Heisenberg equation for the position operator is then given by
\begin{eqnarray}
\frac{d x_i}{d t} &=& i[H,x_i] \nonumber \\
&=& i\left( \left[ \frac{p^2}{2m},x_i\right] + [V,x_i] \right)  \nonumber \\
&=& -\frac{p_i}{m}
\end{eqnarray}
where $m$ is the particle's mass and the last line follows because the potential, $V$, is taken to be a function of neither time nor momentum. This condition on $V$ is restrictive, but it still allows for two potentials which are expected to be important to future work involving this operator, namely $V\propto z$ (a constant force in the $z$-direction) and $V\propto 1/r$ (a Coulomb/Newton force). Notably, the restriction to time-independent potentials here means that the formalism will have to be modified if one wishes to include the effects of a realistic measuring device due to the time-dependent nature of the coupling between a system and its environment. Because of $\vec{x}$ 's simple evolution the commutator of $s$ with the Hamiltonian is not nested. We evaluate it by observing that the only sensible interpretation of a non-power function of an operator is the power series representing it, provided that the series actually converges after acting it term by term on a wavefunction. In light of this we have
\begin{eqnarray}
[H,s] &=& \int_0^td t' \left[H,\sqrt{1-\left(\frac{p}{m}\right)^2}\right] \nonumber \\
&=& \int_0^td t' \left[ H,\sum_{n=0}^\infty {1/2\choose n}\left(\frac{i p}{m}\right)^{2n}\right] \nonumber \\
&=& \sum_{n=0}^\infty {1/2\choose n}\frac{(-1)^n}{m^{2n}}\int_0^td t' [H,p^{2n}]  \nonumber \\
&=& \sum_{n=0}^\infty {1/2\choose n}\frac{(-1)^n}{m^{2n}}\int_0^td t' [V,p^{2n}] \label{eq:hs}
\end{eqnarray}
where ${a\choose b}$ is a binomial coefficient. We can now put the pieces together and write down an expression for the expectation value of the four-distance traveled by a particle in time $t$. Ehrenfest's theorem says of the operator $s$
\begin{equation}
\frac{d}{d t}\langle s \rangle = i\langle [H,s]\rangle + \left\langle \frac{\partial s}{\partial t}\right\rangle.
\end{equation}
Integrating both sides, plugging in~Eq.~\eqref{eq:ds} and~Eq.~\eqref{eq:hs} and using Cauchy's formula on the resulting double integral gives our definition for the expectation value of the quantum distance operator $s$:
\begin{equation}
\langle s\rangle = \int_0^td t' \left\langle \sqrt{1-\left(\frac{p}{m}\right)^2}\right\rangle\  +\ i \sum_{n=0}^\infty {1/2\choose n}\frac{(-1)^n}{m^{2n}}\int_0^td t' (t-t')\langle [V,p^{2n}]\rangle .
\label{eq:s}
\end{equation}
We can hazard a guess at the physical meaning of $\langle s\rangle$ by considering its construction, which began by integrating the four-distance element in Minkowski space. For a classical object this quantity would be the invariant interval of the object's motion. We can thus interpret $\langle s\rangle$, at least qualitatively, as the weighted average of the invariant intervals corresponding to well-defined trajectories that a classical object might take. The $s$-operator is an extended-time observable so if we ever want to measure it it will necessarily involve probing a quantum system repeatedly. However, if we estimate $\langle s\rangle$ using values from a series of projective measurements we are really evaluating how a particle interacts with the detector and not how it moves under the system's Hamiltonian.  In the next section we consider how one might go about measuring the physical quantity corresponding to the new operator $s$.

\section{Measurement Prospects for the $s$-Operator}
\label{sec:measure}
Interpretations in quantum mechanics, almost as a matter of course, are fraught with difficulty. This is especially true when it comes to the various notions of trajectory which enter into the discussion. While a classical particle can be in only one place at one time, wave/particle duality means the same is not true of a quantum particle. This leads to the type of statement made at the beginning of section~\ref{sec:intro}; that quantum particles simply do not follow definite trajectories. Nevertheless, in a variety of instances some mathematical objects which are in some way related to the classical trajectory have been found to be useful. Among the more familiar of these are the paths from Feynman's path integral formulation~\cite{das} and the Bohmian trajectories~\cite{sanz} from the de Broglie-Bohm formulation of quantum mechanics. Much of the time these quantum trajectories are used either as an aid to numerical computation or in the semiclassical regime where classical and quantum degrees of freedom can become coupled. As such, they are often viewed as a crutch for our innately classical mind-set and in that light questions of their reality or interpretation seem unimportant. Although it was not widely appreciated at first, the weak measurement formalism of Aharonov, Albert and Vaidman~\cite{aharonov} provides a means to define and measure experimentally~\cite{matzkin} a \emph{weak quantum trajectory} (hereafter weak trajectory). While reconstructions of Bohmian and weak trajectories from experimental data have been carried out~\cite{albareda}, the calculations in this paper apply only to free particles. As such we cannot yet compare our results to these experiments. Nevertheless, we now briefly discuss both the weak measurement formalism and its use in measuring weak trajectories since they provide an avenue by which our future results might be put to the test.

In the conventional quantum theory, measurements are taken to be both instantaneous and projective. In a great many circumstances these assumptions are thoroughly reasonable and cause no problems. However, it can happen that the system to be measured has dynamics which are fast compared to the measurement process and in such situations it is not appropriate to think of the measurement as being instantaneous. Moreover, if one wants to know about a system's behavior at several moments in time then the destructive (projective) measurement erases useful information. Consider an observable $A$ of a system which is coupled to a measuring device in the von Neumann sense~\cite{von,mello}. This entails modifying the Hamiltonian of the system to include interaction with the measuring device, which is taken to be another quantum object. If we wish to infer the value of $A$ via its interaction with a variable of the measuring device $\xi$ then the interaction Hamiltonian is
\begin{equation}
H_{int} = \lambda f(t)g(\vec{r})A\xi
\label{eq:intham}
\end{equation}
\noindent where $\lambda$ is a coupling constant indicating the strength of the interaction and $f(t)$ is a factor which essentially accounts for the finite duration of the measurement. Its integral over all time must be finite. If such an integral is not equal to unity, we can for convenience absorb its value into the definition of $\lambda$ so that
\begin{equation}
\int_0^\infty f(t)\ dt = 1.
\end{equation}
\noindent Note that the limit of instantaneous measurement corresponds to $f(t)$ becoming a Dirac delta function. The function $g(\vec{r})$ plays a similar role to that of $f(t)$ but for space instead of time. It is a measure of how much the system's and measuring device's wavefunctions overlap. It is, in other words, related to the probability that the measuring device is actually able to ``see" the system. After interacting with the measuring device the system's state is certainly changed by the experience - an unavoidable fact of the quantum world - but if $\lambda$ is small enough the system's state remains practically unchanged. A device possessing small $\lambda$ will be referred to as a weak measuring apparatus (WMA). Of course, there is no free lunch: in exchange for not disturbing the system too badly WMA's are able to extract only small amounts of information from  the system. In practice this reduction of information flow is countered by repeating the experiment many times and/or on large ensembles. After a weak measurement the system continues to evolve according to its propagator $U$. Meanwhile, the mean of the variable conjugate to $\xi$ is shifted by an amount proportional to $\Re \langle A\rangle_w$ where $\langle A\rangle_w$ is the weak value of the measurement of $A$. Suppose the system is prepared in an initial \emph{pre-selected} state $|\psi \rangle$ at $t = t_0$. Then the weak value of $A$ is defined by
\begin{equation}
\langle A\rangle_w = \frac{\langle \chi (t_w)|A|\psi(t_w)\rangle}{\langle \chi(t_w)|\psi(t_w)\rangle}
\label{eq:wv}
\end{equation} 
\noindent where $t_w$ is the time at which the weak measurement occurs and $|\chi \rangle$ is the result of a subsequent projective measurement of some non-$A$ observable. Since it manifests only after the weak part of the measurement happens, say at a time $t_f > t_w$, $|\chi \rangle$ is known as a \emph{post-selected} state. In their definitions neither $|\psi \rangle$ nor $|\chi \rangle$ are given at $t_w$, as required by~Eq.~\eqref{eq:wv}, so we should understand $|\psi(t_w)\rangle$ and $|\chi(t_w)\rangle$ to be the forward and backward propagated states, respectively:
\begin{eqnarray}
|\psi(t_w)\rangle &=& U(t_w,t_0)|\psi(t_0)\rangle, \\
\langle \chi(t_w)| &=& \langle \chi(t_f)|U^\dagger(t_f,t_w).
\end{eqnarray}
Now that we have explored - however briefly - the concept of a weak quantum measurement we follow Matzkin~\cite{matzkin,matzkin12} and define the system's weak trajectory. Suppose that the system is to interact with $N$ identical WMA's that seek to measure the system's position $\vec{r}$, each of which having a wavefunction tightly localized at some point $\vec{R}_k$ for $k = 1,...,N$. It is often mathematically preferable and physically reasonable to take the WMA's wavefunctions to be Gaussians centered on each of the $\vec{R}_k$. In such a case the interaction Hamiltonian between the system and the $k^{th}$ WMA takes the form
\begin{equation}
H_{int,k} = \lambda f(t)g(|\vec{r}-\vec{R}_k|^2)\vec{r}\cdot \vec{R}_k.
\end{equation}
Provided that the interaction time (width of $f(t)$) is sufficiently short compared to the system's dynamics, we can take it to occur instantaneously and the weak value of position measured by the $k^{th}$ WMA is~\cite{matzkin12}
\begin{equation}
\langle \vec{r}(t_k) \rangle_w = \frac{\langle \chi(t_k)|\vec{r}g(|\vec{r}-\vec{R}_k|^2)|\psi(t_k)\rangle}{\langle \chi(t_k)|\psi(t_k)\rangle}
\end{equation}
\noindent where $t_k$ is the center of the $k^{th}$ WMA's duration function $f(t)$. If one labels the WMA's according to the order in which they interact with the system (\emph{i.e.} the $k=1$ WMA interacts with the system first, the $k=2$ one interacts second, etc.) then the weak trajectory corresponding to the pre-selected state $|\psi\rangle$ and post-selected state $|\chi\rangle$ is defined by the set
\begin{equation}
WT_{\psi \chi} = \{(t_k,\ \Re \langle \vec{r}(t_k)\rangle_w)|\ k=1,...,N\}.
\end{equation}
From the weak trajectory approximations to the invariant four-distance are readily calculated. Take for granted that the particle is moving through Minkowski spacetime. Then the time-component of the classical interval is unity and the space components are related to the weak measurements of position. If the $t_k$ are equally spaced so that $t_k - t_{k-1} = \Delta t$ for all $k$ then we can approximate the derivative
\begin{equation}
\frac{d\vec{x}}{dt}\Big|_{t = t_k} \approx \frac{\Re \langle \vec{r}(t_k)\rangle_w - \Re \langle \vec{r}(t_k-1)\rangle_w}{\Delta t}.
\end{equation}
Now we can write an expression for the expectation value of the distance operator
\begin{align}
\langle s \rangle &= \int_0^t dt'\ \sqrt{1 - \frac{\langle \vec{x}\rangle}{dt'}\cdot \frac{\langle \vec{x}\rangle}{dt'}} \\
&\approx \Re \sum_{k=1}^N \sqrt{1 - \frac{1}{(\Delta t)^2}(\langle \vec{r}(t_k)\rangle_w^2 - 2\langle \vec{r}(t_k)\rangle_w\langle \vec{r}(t_{k-1})\rangle_w + \langle \vec{r}(t_{k-1})\rangle_w^2)}
\end{align}
in terms of (weak) experimental measurements. While series of weak measurements have been made in actual quantum systems the experimental set-ups involved are substantially more complicated than the free particle. Even a relatively simple interference experiment confines the particle being experimented on to a finite region of space. The non-trivial geometry (reduction of symmetry) forces one to write the Hamiltonian in terms of things like step- and/or delta-functions which make the expression for $\langle s\rangle$ unwieldy. And that's to say nothing of the contribution from von Neumann's interaction Hamiltonians, Eq.~\eqref{eq:intham}.

Interaction and finite geometry may make the actual calculation of $\langle s \rangle$ more difficult, but they don't change the overall scheme for calculation, measurement and interpretation. Relativity, on the other hand, topples this whole section or at least means it needs adjustment. For instance, the definition of the weak trajectory given above relies on a fixed order of events (interactions with the WMA's) with definite intervals of time separating them. This clearly needs modification when a change of frame can shuffle the order of events. Even if all the intervals between detections are spacelike, the definition of weak measurement refers to single-particle wavefunctions which don't always represent probability amplitudes in relativistic quantum theory. Because of the decision to use Heisenberg's equation, instead of a relativistic analog thereof, in the construction of section~\ref{sec:der} we already knew that the $s$-operator would need to be reworked in order to respect Einstein's theory. Now we can see that our proposal for measuring $\langle s\rangle$ will similarly need revision if it is to be applied to particles of appreciable momentum, $p \sim m$.

To summarize, because it is time-extended experimental values of $\langle s\rangle$ must be formed from measurements taken at multiple near-by times. Since they are nondestructive, weak measurements provide a way this could be achieved. The computation of $\langle s\rangle$ for a weak measurement setup is possible but it will be quite the undertaking. To give us confidence that such effort will not be wasted we turn now to simpler calculations. In particular, the rest of this paper deals with an initially well-localized free particle and its expected value of $s$.

\section{Quantum Distance Traveled by a Free Particle}
\label{sec:free}
For the free particle the second term in~Eq.~\eqref{eq:s} vanishes and we have
\begin{equation}
\langle s\rangle = \int_0^td t' \left\langle \sqrt{1-\left(\frac{p}{m}\right)^2}\right\rangle.
\label{eq:sfree}
\end{equation}
Since $[H,s]=0$ in this case the eigenstates of $H=p^2/2m$ are also eigenstates of $s$. Because they have definite momenta one sees that plane waves
\begin{equation}
\langle x|\psi \rangle \propto \exp(-i Et-i \vec{p}\cdot \vec{x})
\end{equation}
have for their $s$-eigenvalues
\begin{equation}
t\sqrt{1-\left(\frac{p}{m}\right)^2}
\label{eq:eig}
\end{equation}
in which $p$ represents a momentum eigenvalue as opposed to the momentum operator. This shows that completely delocalized planewave states travel a continuum of distances which depend, in the classical way, on their momenta. On the other hand, a planewave particle-in-a-box has a discrete spectrum of momentum eigenvalues and thus, according to~Eq.~\eqref{eq:eig}, can only be found to have traveled certain distances within the box. This implies an \textit{effective} discretization of space or time or both when it comes to the motion of a confined  particle. It is important to emphasize that~Eq.~\eqref{eq:eig} has nothing at all to say about the structure of spacetime, meaning that the discretization phenomenon is limited to the values of $s$ which may be observed and has no baring on the geometry inside the box.

The case of planewaves was easy to analyze but it doesn't do much for our goal of testing the equivalence principle in the semiclassical regime since the totally delocalized planewave states lack any classical analog. We must therefore turn our attention to localized wavepackets for which the classical limit corresponds to particles of definite position and momentum. To facilitate the calculation we first notice that computing~Eq.~\eqref{eq:sfree} by power series expansion will involve computing all the even moments of the wavepacket, $\langle p^{2n}\rangle$. Let $U$ be the unitary propagator so that
\begin{equation}
|\psi(t)\rangle = U|\psi(0)\rangle
\end{equation}
for any initial state ket $|\psi(0)\rangle$. For $V = 0$ the propagator depends only on time so that $[U,p]=0$, allowing us to ignore the time evolution of the initial wavefunction while calculating $\langle p^{2n} \rangle$:
\begin{eqnarray}
\langle \psi(t)|p^{2n}|\psi(t) \rangle &=& \langle \psi(0)|U^\dagger p^{2n}U|\psi(0)\rangle \nonumber \\
&=& \langle \psi(0)|p^{2n}U^\dagger U|\psi(0)\rangle \nonumber \\
&=& \langle \psi(0)| p^{2n}|\psi(0)\rangle. 
\end{eqnarray}
We remark in passing that this simplification will not be possible for particles subject to linear and Coulombic potentials since the corresponding propagators have position dependence and so fail to commute with momentum operators. The wavepacket we choose to analyze saturates the uncertainty bound: a Gaussian initially centered on the origin with initial mean momentum in the negative $z$-direction of $p_{0}$. The initial wavefunction in momentum space is
\begin{equation}
\langle p|\psi(0)\rangle = \frac{\sqrt{2}\sigma^{3/2}}{\pi^{1/4}}\exp \left[-\frac{\sigma^2}{2}(\vec{p}+p_{0}\hat{p}_z)^2\right]
\end{equation}
where $\sigma$ is the initial spread of the wavefunction in position space. The even moments are then
\begin{eqnarray}
\langle p^{2n}\rangle &=& \frac{2\sigma^3}{\sqrt{\pi}}\exp(-\sigma^2 p_{0}^2)\int_0^\infty d p\ p^{2n+2} \exp(-\sigma^2p^2)\int_{-1}^1d (\cos\theta)\exp(-2\sigma^2pp_{0}\cos\theta) \nonumber \\
 &=&  -\frac{2\sigma^3}{\sqrt{\pi}p_{0}}\exp(-\sigma^2 p_{0}^2)\int_0^\infty dp\ p^{2n+1} \exp(-\sigma^2p^2)\sinh(2\sigma^2pp_{0}).
\label{eq:emom}
\end{eqnarray}
To evaluate this last integral we make frequent use of identities and formulas from Buchholtz' compendium on confluent hypergeometric functions~\cite{conf} and begin with a change of variables $u=(\sigma p)^2$. Then we write the hyperbolic sine as a $_{0}F_{1}$ generalized hypergeometric function so that
\begin{eqnarray}
 \int_0^\infty d p\ p^{2n+1} \exp(&-&\sigma^2p^2)\sinh(2\sigma^2pp_{0}) \nonumber \\ 
&=& \frac{\sqrt{\pi}p_{0}}{2\Gamma(3/2)\sigma^{2n+1}}\int_0^\infty d u\ e^{-u}u^{n+1/2}\ _{0}F_{1}(\ , 3/2, \sigma^2p_{0}^2u)
\end{eqnarray}
where $\Gamma(z)$ is the gamma function. This integral is the special case of the integral representation
\begin{equation}
_{1}F_{1}(a,b,z) = \frac{1}{\Gamma(a)}\int_0^\infty d t\ e^{-t}t^{a-1}\ _{0}F_{1}(\ ,b,zt),
\end{equation}
which is valid so long as the real part of $a$ is positive, with $a=n+3/2$, $b=3/2$ and $z = \sigma^2p_{0}^2$. In this way the integral in $\langle p^{2n}\rangle$ can be expressed as a $_{1}F_{1}$ hypergeometric function which, after an application of Kummer's transformation, reduces to a generalized Laguerre polynomial. Thus, the even moments of a free Gaussian wavefunction with average momentum $-p_{0}\hat{p}_z$ are
\begin{equation}
\langle p^{2n}\rangle = -\frac{1}{2\sqrt{\pi}}\left(\frac{-1}{2\sigma}\right)^{2n}\Gamma(-n-1/2)\Gamma(2n+2)L_n^{1/2}(-\sigma^2 p_{0}^2)
\label{eq:mom}
\end{equation}
where $L_n^{\alpha}(z)$ is a generalized Laguerre polynomial. To evaluate~Eq.~\eqref{eq:sfree} we expand the radical in a power series, plug in 
the moments~Eq.~\eqref{eq:mom} and simplify the resulting combination of gamma functions. Since the expectation values are in this case time-independent the time integration in~Eq.~\eqref{eq:sfree} is trivial and we have
\begin{equation}
\langle s\rangle = -\frac{t}{2\sqrt{\pi}}\sum_{n=0}^\infty \Gamma(n-1/2)x^nL_n^{1/2}(-\beta^2/x)
\label{eq:exps}
\end{equation}
where we have introduced the parameter $x \equiv 1/(m\sigma)^2$ for convenience (note that it is proportional to $\hbar^2$) and $\beta$ is the usual relative velocity of the particle, $\beta = v_{0}/c$. The expectation value in~Eq.~\eqref{eq:exps} is clearly dependent on the mass of the particle whose motion it describes, implying that UFF does not hold on the quantum scale. To validate these results we take the classical $\hbar \to 0$ limit of~Eq.~\eqref{eq:exps}. This is achieved by replacing the Laguerre polynomials with the first term in their asymptotic expansions,
\begin{equation}
L_n^\alpha(z) \sim \frac{(-z)^n}{\Gamma(n+1)},
\end{equation}
and summing the resulting series:
\begin{eqnarray}
\lim_{\hbar \to 0} \langle s\rangle &=& -\frac{t}{2\sqrt{\pi}}\sum_{n=0}^\infty \frac{\Gamma(n-1/2)}{\Gamma(n+1)}\beta^{2n} \nonumber \\
&=& t\sqrt{1-\beta^2}.
\label{eq:lim}
\end{eqnarray}
This is the classical result for the four-interval traversed by a free particle with mass $m$ and constant momentum $p_{0}$. We consider equations~Eq.~\eqref{eq:exps} and~Eq.~\eqref{eq:lim} to be proof that the distance operator introduced in this paper is interesting (or at least self-consistent) and worthy of study. In particular, we see that the interval's expectation value depends on the particle's mass on the scale set by $\hbar$ but in the classical limit the particle moves \emph{independent of its own properties}. We take this to indicate that at least the UFF version of WEP is only an approximate feature of the quantum world. It then seems plausible that the $s$-operator could be used with a linear gravitational potential to assess the validity of NEP in a quantum context, but this is beyond the scope of the current work.

While the result of taking the classical limit of the expectation value of the distance operator $s$ for a localized wavepacket,~Eq.~\eqref{eq:lim}, is compelling there is a complication. Having a gamma function in the numerator of the terms in~Eq.~\eqref{eq:exps} without one or more such functions in the denominator does not bode well for the series's convergence. Indeed, since~\cite{asy}
\begin{equation}
\lim_{n\to \infty} \frac{L_{n+1}^\alpha(z)}{L_n^\alpha(z)} = 1,
\end{equation}
a quick check with the ratio test shows that the series diverges everywhere \emph{except} in the classical limit. What are we to make of this? Retracing our steps we find that this divergence is the result of combining the relativistic interval~Eq.~\eqref{eq:mink} with the \emph{non}-relativistic Heisenberg equation. One of the first things we did was to expand the radical in Eq.~\eqref{eq:mink} as a formal power series, but the series for $\sqrt{1-x^2}$ diverges if $|x|>1$. This does not actually invalidate Eq.~\eqref{eq:lim} because the $\hbar \to 0$ limit is not distinguishable in this case from the $m\to \infty$ limit. In the latter case, as $m$ gets bigger numbers $p$ from a larger portion of the real line satisfy $|p/m|^2 < 1$. When $m$ goes all the way to infinity the binomial series converges for all momenta so the exchange of summation and integration implicitly made in Eq.~\eqref{eq:exps} becomes legitimate. It seems then that the analysis leading to the classical limit of $\langle s\rangle$ is \emph{only} valid in that limit and we must resort to numerical methods (or at least put away the binomial series) to evaluate $\langle s\rangle$ in the quantum and semi-classical regimes. If Eq.~\ref{eq:lim} applies only to classical particles, which necessarily have $p \ll m$, why go through all the trouble of solving the general integral in Eq.~\ref{eq:emom} and resumming the infinite series? Wouldn't a first order approximation to $\sqrt{1-x^2}$ work just as well? Certainly, for the purposes of computing a numerical value of $\langle s\rangle$, it makes virtually no difference whether one takes two terms or a thousand in the binomial expansion. The benefit to keeping all the terms is that the classical limit, Eq.~\ref{eq:lim}, is seen to agree with the classical result to \emph{all orders}.

\begin{figure}[h!]
\centering
\includegraphics[scale=0.6]{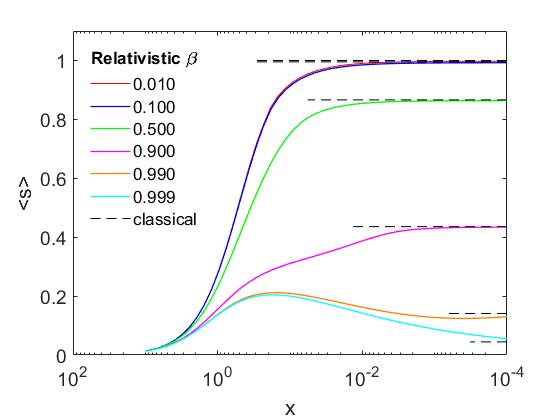}
\caption{Numerical integration of the truncated integral~Eq.~\eqref{eq:tint}. The trivial time integral has been carried out and, as it only serves to set the overall scale of the graph, $t$ is set to unity. Each curve corresponds to a different value of $\beta$, with the $\beta = 0.01$ and $\beta = 0.1$ curves nearly coinciding. For reference the classical interval values, $\sqrt{1-\beta^2}$, are displayed as horizontal asymptotes of the curves.}
\label{fig:int}
\end{figure}

Numerical analysis is facilitated by expressing $\langle s\rangle$ as an integral over dimensionless parameters. Undoing the time integral in~Eq.~\eqref{eq:sfree} and expressing $\langle s\rangle$ in momentum space gives
\begin{equation}
\frac{d\langle s\rangle}{d t} = \int d^3\vec{p}\ |\langle \psi | p\rangle|^2\sqrt{1-\left(\frac{p}{m}\right)^2}.
\label{eq:sint}
\end{equation}
We can manually enforce special relativity and make numerical evaluation of the integral~Eq.~\eqref{eq:sint} easier in one stroke. Recall that $\langle p|\psi  \rangle$ is a Gaussian centered on $p_{0}$. If we assume that $p_{0} \ll m$ and that $\sigma$ is not too large then the integrand is negligibly small in the relativistic $p  \gtrsim m$ regime. Thus, it is reasonable to approximate the infinite integral in~Eq.~\eqref{eq:sint} by a truncated finite integral. Carrying out the integral over solid angle in momentum space exactly, separating the resulting hyperbolic sine function into exponentials, changing variables to $\rho = \sigma p$ and applying the truncation gives
\begin{align} 
 \frac{d\langle s\rangle }{d t} \approx \frac{1}{\beta}\sqrt{\frac{x}{\pi}}\biggl\{ \int_{\xi_1}^{\xi_2}d \rho &  \exp[-(\rho - \rho_{0})^2]\sqrt{1-x\rho^2}\rho \nonumber \\
&- \int_0^{\xi_3}d\rho  \exp[-(\rho + \rho_{0})^2]\sqrt{1-x\rho^2}\rho \biggr\}, \nonumber \\
\xi_1 = \min(&\rho_{0}/\beta,\max(0,\rho_{0}-3/\sqrt{2})),\nonumber \\
\xi_2 = \min(&\rho_{0}/\beta,\rho_{0}+3/\sqrt{2}), \nonumber \\
\xi_3 = \min(&\rho_{0}/\beta,\max(0,-\rho_{0}+3/\sqrt{2})) \label{eq:tint}
\end{align} 
in which $\rho_{0} = m\sigma v_{0} = \beta/\sqrt{x}$. The finite limits of integration are chosen so as to keep the integration variable within three standard deviations of each Gaussian as well as respecting relativity. Figure~\ref{fig:int} shows the result of numerical integration of~Eq.~\eqref{eq:tint} for a range of $x$-values and initial particle mean velocities.
The graph is set up so that the parameter $x$ decreases to the right which means that side of the chart shows the classical limit. We see that the curves, each representing the expected interval of a particle with relative velocity $\beta$, are indeed approaching their classical values when $x$ gets small, as~Eq.~\eqref{eq:lim} insists upon. On the large-$x$ side of the graph all of the intervals become light-like. This makes sense when one recalls that $x$ goes like $1/m^2$: massless particles in relativity must traverse light-like intervals. While most of the curves climb from nearly zero to their asymptotic values monotonically, the curves corresponding to the fastest particles considered here ($\beta = 0.990$ and $\beta = 0.999$) achieve maxima which are actually larger than the classical values. The physical meaning of these curves is not clear. Since the operator they're based on is a combination of relativistic (Minkowski metric) and non-relativistic (Heisenberg's equation) components, it is also not clear whether the high-$\beta$ curves are even valid. If that's the case then why display these curves at all in Figure~\ref{fig:int}? In part it is to show that the macroscopic limit of the $s$-operator is consistent with its non-quantum analog even at momenta which should invalidate it. The main reason, though, for computing and displaying values of $\langle s\rangle$ for which the particle has $\langle p \rangle \sim m$ is for comparison with future results. There are several conceivable generalizations of $s$ to the arena of quantum field theory and we would like to see if/how they smoothly tie into the semi-relativistic operator of this paper.

\begin{figure}[h!]
\centering
\includegraphics[scale=0.6]{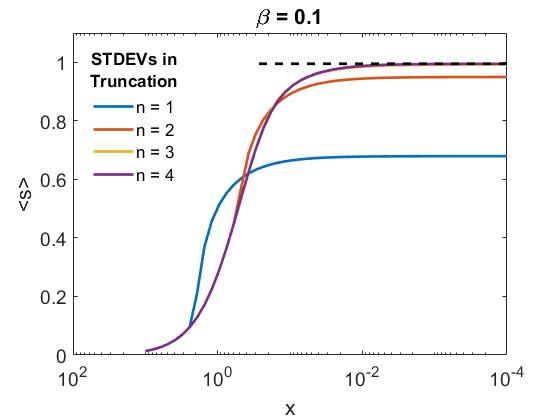}
\caption{Expectation values of \emph{s} approximated by~Eq.~\eqref{eq:tint} modified to have a variable truncation range. The curves each represent integrals whose limits of integration are $\pm n$ standard deviations away from the center of the Gaussian in the integrand. The curves with $n=3$ and $n=4$ coincide, indicating that we achieve sufficient numerical accuracy with $n=3$.}
\label{fig:stdev}
\end{figure}

The decision to truncate the infinite integrals in~Eq.~\eqref{eq:tint} at three standard deviations is arbitrary, but numerical experimentation shows that it makes little difference if one extends the region of integration further, provided that $p$ gets no greater than $m$. Figure~\ref{fig:stdev} shows an example for $\beta = 0.1$.

\section{Conclusions \& Future Work}
\label{sec:fut}
We have presented a new quantum mechanical operator based on the four-interval element in Minkowski space. Using Heisenberg's equation for the time evolution of operators we deduced its expectation value and found that it depends in a complicated way on the potential the particle is exposed to. The free particle ($V=0$) case was then analyzed for two types of states: completely delocalized planewaves and maximally localized Gaussian wavefunctions. The classical limit of $\langle s\rangle$ for the latter states was then shown to agree with the standard classical result for a particle moving at constant speed $v_{0}$. Since it seems likely that $\langle s\rangle$ is, in some sense, an average over the possible paths the particle could take, it provides us with information on whether and how a quantum particle's mass influences its mean trajectory. The truncated integral~Eq.~\eqref{eq:tint} and the classical limit~Eq.~\eqref{eq:lim} then imply that the UFF version of WEP fails to hold for quantum particles but is restored for classical masses and momenta. This result is in qualitative agreement with quantum time-of-flight considerations~\cite{ali,viola,davies}.  The Minkowski interval element was quantized in a non-rigorous way, making the operator and calculations based on it highly speculative. However, it has been demonstrated that the $s$ operator is self-consistent and useful.

There are a number of possible extensions to this work which may prove insightful. Some of them are, in no particular order:
\begin{itemize}
\item Effects of confinement. We have already addressed this for particle-in-a-box eigenstates, but what about a localized particle-in-a-box? While the $s$ operator itself is the same in this case as it is for the free particle in infinite space, the computation of its expectation values is complicated by the requirement that the wavefunction vanish at the box walls.
\item Having established that use of the new operator gives sensible results in the simplest case, its expectation value for a localized particle in either linear or Newton/Coulomb inverse potentials should be calculated to bring gravity into the picture.
\item What effect does bestowing a particle with orbital and/or intrinsic angular momentum have on the four-interval it traverses in a given time?
\item The $s$ operator as defined in this work is an amalgamation of relativistic and non-relativistic parts. Can this be remedied by replacing the use of Heisenberg's equation with an analogous one based on the Dirac or Klein-Gordon equations? What happens if we use a Halimtonian containing relativistic correction terms?
\item Some experiments with optical tweezers have made repeated position measurements of an individual trapped particle. The potential of a trapped particle is complicated, but in certain cases the harmonic oscillator is good approximation to it. This makes the $s$ operator for the harmonic potential $V \propto x^2$ interesting as there may be existing experimental data available with which to test it.
\item How do the interactions of several particles effect the distances they travel? What sort of interval is covered by a Schr\"{o}dinger cat state? By an entangled pair?
\end{itemize}

The investigation of these and other topics are the subject of ongoing work by the author.

\bibliographystyle{unsrt}
\bibliography{quantized_distance_bib}

\end{document}